\documentclass[fleqn,usenatbib]{mnras}

\usepackage[T1]{fontenc}

\DeclareRobustCommand{\VAN}[3]{#2}
\let\VANthebibliography\thebibliography
\def\thebibliography{\DeclareRobustCommand{\VAN}[3]{##3}\VANthebibliography}

\usepackage{graphicx}	% Including figure files
\usepackage{amsmath}	% Advanced maths commands
\usepackage{amssymb}	% Extra maths symbols

\usepackage{gensymb}

\def \aj {AJ}
\def \mnras {MNRAS}
\def \apj {ApJ}
\def \apjs {ApJS}
\def \apjl {ApJL}
\def \aap {A\&A}

\defcitealias{fermipaper2018}{SA2018}

\usepackage{newtxtext,newtxmath}
\title[The radio counterpart of LS 2355]{Particle acceleration at the bow shock of runaway star LS 2355:\\ non-thermal radio emission but no $\gamma$-ray counterpart}

\author[J. van den Eijnden et al.]{J. van den Eijnden,$^{1}$ S. Mohamed,$^{2}$ F. Carotenuto,$^{3}$ S. Motta,$^{4}$ P. Saikia,$^{5}$ D. R. A. Williams-Baldwin$^{6}$
\\
$^{1}$Department of Physics, University of Warwick, Coventry CV4 7AL, UK\\
$^{2}$ University Of Virginia, Astronomy Building, 530 McCormick Road, Charlottesville, VA 22904, USA\\
$^{3}$ Astrophysics, Department of Physics, University of Oxford, Keble Road, Oxford OX1 3RH, UK\\
$^{4}$ Istituto Nazionale di Astrofisica, Osservatorio Astronomico di Brera, via E. Bianchi 46, I-23807 Merate (LC), Italy\\
$^{5}$ Center for Astrophysics and Space Science (CASS), New York University Abu Dhabi, PO Box 129188, Abu Dhabi, UAE \\
$^{6}$ Jodrell Bank Centre for Astrophysics, School of Physics and Astronomy, University of Manchester, Manchester M13 9PL, UK \\
}

\date{Accepted XXX. Received YYY; in original form ZZZ}

\pubyear{2023}

\begin{document}
\label{firstpage}
\pagerange{\pageref{firstpage}--\pageref{lastpage}}
\maketitle

\begin{abstract}
Massive stars that travel at supersonic speeds can create bow shocks as their stellar winds interact with the surrounding interstellar medium. These bow shocks -- prominent sites for mechanical feedback of individual massive stars -- are predominantly observed in the infrared band. Confirmed high-energy emission from stellar bow shocks has remained elusive and confirmed radio counterparts, while rising in recent years, remain rare. Here, we present an in-depth multi-wavelength exploration of the bow shock driven by LS 2355, focusing on its non-thermal properties. Using the most-recent \textit{Fermi} source catalogue, we rule out its previously-proposed association with an unidentified $\gamma$-ray source. Furthermore, we use deep ASKAP observations from the Rapid ASKAP Continuum Survey and the Evolutionary Map of the Universe survey to identify a non-thermal radio counterpart: the third spectrally confirmed non-thermal bow shock counterpart after BD +43$\degree$ 3654 and BD +60$\degree$ 2522. We finally use \textit{WISE} IR data and \textit{Gaia} to study the surrounding ISM and update the motion of LS 2355. Specifically, we derive a substantially reduced stellar velocity, {\color{black} $v_* = 7.0\pm2.5$ km/s}, compared to previous estimates. The observed non-thermal properties of the bow shock can be explained by an interaction between the wind of LS 2355 and a dense HII region, at a magnetic field close to the maximum magnetic field strength allowed by the compressibility of the ISM. Similar to earlier works, we find that the thermal radio emission of the shocked ISM is likely to be substantially suppressed for it to be consistent with the observed radio spectrum.

\end{abstract}

\begin{keywords}
shock waves; stars: early-type; stars: individual: LS 2355; radio continuum: general; acceleration of particles; gamma-rays: general
\end{keywords}

\section{Introduction}

Through mass loss in the form of powerful stellar winds, massive stars can greatly impact their interstellar surroundings. Such massive stellar feedback, for instance taking the form of wind-blown nebulae around Wolf-Rayet stars \citep[e.g.,][]{Prajapati2019}, deposits significant amounts of energy and momentum into the interstellar medium (ISM) that can heat, shape, and energize these surroundings. As these feedback processes are often associated with shock formation at the interaction site between the wind and ISM \citep[e.g.,][]{delpalacio2018}, or between stellar winds in e.g., colliding wind binaries \citep[e.g.,][]{reimer2006}, the stellar wind energy budget may power the acceleration of charged particles into a non-thermal particle population \citep{delvalle2012,delvalle2018,martinez2023}. Particle-accelerating feedback structures around massive stars may then show up at either high \citep[X-rays, gamma-rays; e.g.,][]{debecker2017} or low (radio) frequency observations \citep[e.g.,][]{benaglia2010,benaglia2021}: such non-thermal emission can be dominated by synchrotron emission from the accelerated population subject to the shock’s magnetic field, inverse Compton emission as this population interacts with either the stellar or (shocked) ISM photon field, or $\gamma$-ray emission from hadronic interactions.

Sites of massive (non-cataclysmic) stellar feedback are often found in runaway massive stars that move supersonically through the ISM. Ejected from their birth location via either dynamical interactions or the supernova of a binary companion \citep{blaauw1961,poveda1967}, the runaway launches a stellar wind that creates a bow shock in the ISM in the star’s direction of motion. Galactic runaway massive stars with bow shocks are typically found in or close to the Galactic plane through the bright infrared emission from swept up dust heated by the massive star’s radiation. Recent catalogues of such infrared-selected (candidate) bow shocks include the E-BOSS catalogues \citep{peri2012,peri2015}, the 709 systems identified by \citet{kobulnicky2016} and the 453 systems identified by the Milky Way Project \citep{mw_project}, 311 of which did not appear in the former catalogue. Significantly fewer systems are known at other wavelengths: in optical bands, for instance, absorption in the Galactic plane or close-by dense regions often prevents the detection of an optical (continuum or emission line) counterpart \citep{brown2005,meyer2016}.

At both ends of the electromagnetic spectrum where non-thermal evidence of particle acceleration may show up, bow shock counterparts are even more rare -- thereby complicating the observational characterization of the shocked stellar wind and, in particular, the particle acceleration process. In the radio band, nine (candidate) bow shocks have been identified \citep{benaglia2010,benaglia2021,moutzouri2022,vandeneijnden2022_velx1,vandeneijnden2022_racs}. Notably, with the exception of two of those seen with the VLA \citep{benaglia2010,moutzouri2022}, these (candidate) counterparts have been identified using the new MeerKAT and Australian Square Kilometre Array Pathfinder (ASKAP) telescopes in the past two years: the exceptional sensitivity of these arrays to extended structures of low surface brightness has proved crucial in efficiently detecting bow shock radio emission. Of these nine systems, only two -- BD+43$\degree$ 3654 and BD+60$\degree$ 2522 -- show direct observational evidence for particle acceleration through the presence of non-thermal radio spectral signatures in parts of the bow shock \citep{moutzouri2022}. In all others, a lack of spectral information prevents such a direct observational identification of non-thermal emission, and a significant or dominant contribution of thermal (free-free) radio emission from the shocked ISM may be present \citep[][see \citealt{martinez2023} for a recent discussion in distinguishing these emission mechanisms using a simulation perspective]{vandeneijnden2022_velx1}.

At the other end of the spectrum, in the X-ray band, no unambiguously identified bow shock counterparts are known. Only a single marginal detection of non-thermal X-ray emission from a bow shock around a runaway star has been reported to date. This detection was claimed by \citet{lopezsantiago2012} for AE Aurigae (HIP 24575) using \textit{XMM-Newton} observations. However, more recently, \citet{rangelov2019} used new, sub-arcsecond-resolution Chandra data, confirming the presence of the \textit{XMM-Newton} source but also finding that it is neither extended nor coincident with the bow shock’s infrared arc. Instead, the originally proposed X-ray counterpart was strongly suggested to be a background AGN. No other claims of X-ray stellar bow shock detections have been made.

In $\gamma$-rays, \citet[][hereafter {\color{blue}SA2018}]{fermipaper2018} reported the association between two unidentified \textit{Fermi} point sources and two massive stellar IR bow shocks, driven by the massive stars $\lambda$ Cep and LS 2355. Their spatial overlap with the \textit{Fermi} source position uncertainty was, in both cases, used to argue for the association, where the authors note that the bow shock is the object within the error region that is most likely to accelerate particles. Using the non-thermal bow shock emission model by \citet{debecker2017}, \citetalias{fermipaper2018} assessed whether these $\gamma$-ray SEDs can be reconciled with the expected properties of the shock, thereby deriving several of its properties: the maximum energy of electrons, the slope of their number density distribution, and the shock's magnetic field.

Out of these two objects, LS 2355 (also known as HD 99897 and HIP 56021), was not previously known in the aforementioned catalogues of bow shock candidates. Optical and infrared observations of its surroundings indicate that its bow shock is located at the edge of a larger scale HII region, GAL 293.60$-$01.28 \citep[e.g.,][]{georgelin2000,cersomiso2009,lee2012}, that the massive runaway star's wind appears to collide with. Modeling its \textit{Fermi} spectrum, \citetalias{fermipaper2018} found that a low magnetic field of $0.1$~$\mu$G and a modest maximum electron energy of $90$--$125$ GeV can explain the presence of very-high energy emission through inverse Compton scattering of infrared dust emission by the relativistic electron population. However, \citetalias{fermipaper2018} also noted that, to match the normalization of the $\gamma$-ray spectrum, an energy budget for particle acceleration may be required that exceeds the assumed stellar wind kinetic power budget. This potential complication was ascribed to the unknown true parameters of the system -- such as the wind power and the efficiency of convection of relativistic particles. While the $\gamma$-ray source would, if indeed the counterpart, uniquely constrain the properties of the electron population, it alone cannot further unravel this potential discrepancy between the inverse Compton scenario and the stellar wind properties.

Radio observations provide an alternative and complementary constraint on the particle acceleration process; either through direct detection of synchrotron emission from the accelerated population \citep[e.g.,][]{delvalle2012} or via upper limits on this emission \citep[e.g.,][]{debecker2017}. In particular, for a bow shock with detections of both synchrotron and inverse Compton emission, or strong limits on the former, their relative luminosities offer an independent constraint on the strength of the magnetic field and, in turn, the required power budget in the stellar wind \citep{vandeneijnden2022_velx1}. Furthermore, multi-band radio measurements may constrain the non-thermal radio spectrum, which is directly related to the properties of the particle energy spectrum. 

Using data from the Sydney University Molonglo Sky Survey \citep[SUMSS;][]{beck1999}, \citetalias{fermipaper2018} noted that the larger-scale HII region has a radio counterpart. As expected, the integrated flux density of the full region, dominated by its thermal radio emission, greatly exceeds what could feasibly be emitted by the bow shock alone. The advent of Southern pathfinder telescopes to the Square Kilometre Array (SKA), including the Australian SKA Pathfinder (ASKAP), provides a new opportunity to search for a non-thermal radio counterpart of this bow shock. In particular, observations for ongoing Rapid ASKAP Continuum Survey  \citep[RACS;][]{mcconnell2020} and the Evolutionary Map of the Universe survey \citep[EMU;][]{norris2011,norris2021} provide significant improvements in sensitivity at low radio frequencies (UHF, L, and S bands). Their spatial resolution is higher too, but not so high as to resolve out any large-scale, diffuse structures such as radio bow shocks \citep{vandeneijnden2022_racs}. That unique combination of resolution and sensitivity warrants a new search for the non-thermal counterpart of the LS 2355 bow shock.

In this paper, we explore the radio properties of the field around LS 2355 using EMU, RACS, and SUMSS observations\footnote{None of these surveys covered the Northern position of $\lambda$ Cep. We therefore do not further discuss this second source from \citetalias{fermipaper2018} here.}. We furthermore update the search for a $\gamma$-ray counterpart of the bow shock by \citetalias{fermipaper2018}, including the most recent \textit{Fermi} data release. In addition, we include up-to-date proper-motion measurements of LS 2355 by \textit{Gaia} and \textit{WISE} IR data in our analysis. Combining these multi-wavelength data sources, we report the discovery of a non-thermal radio counterpart of the LS 2355 bow shock. Using the non-thermal radio properties and the enhanced, updated $\gamma$-ray position, we can furthermore rule out the proposed association between the $\gamma$-ray source and the bow shock.

\section{Data}
\label{sec:data}

\begin{figure*}
\includegraphics[width=\textwidth]{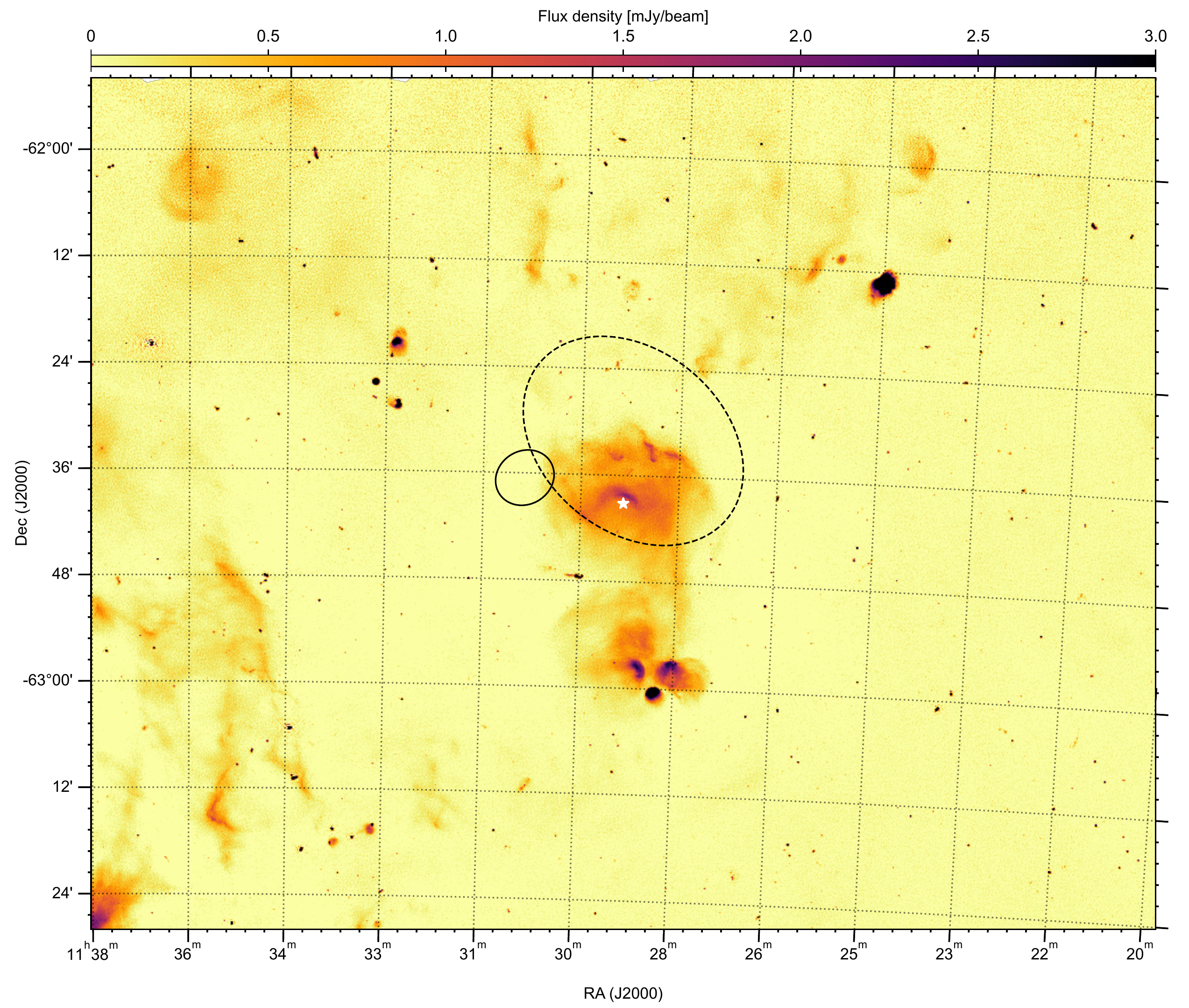}
 \caption{A large-scale cut out of the EMU field covering the position of LS 2355. The field covers $2\times1.6$ degrees, centred at the white star that indicates LS 2355. The Galactic plane stretches along the bottom-left, top-right diagonal direction. The central region covering the runaway star, its bow shock, and the GAL 293.60$-$01.28 HII region, are relatively unaffected by large-scale imaging artefacts. The EMU synthesized beam is shown in bottom left of the zoomed image in Figure \ref{fig:emu_zoom}. The dashed ellipse indicates the 90\% positional uncertainty for the potential \textit{Fermi} counterpart 3FGL J1128.7$-$6232 proposed by \citetalias[][]{fermipaper2018}; the smaller solid ellipse indicated the 90\% positional uncertainty for the corresponding source in the 4FGL-DR4 catalogue 4FGL J1130.5$-$6236c.}
 \label{fig:emu_full}
\end{figure*}

For this multi-wavelength study, we employ publicly-available survey data across radio, IR, optical, and $\gamma$-rays. In the radio band, we employ data from two telescopes. Firstly, the Molonglo Observatory Synthesis Telescope (MOST) consists of two co-linear cylindrical paraboloids oriented in the East-West direction. It has a continuous uv-plane coverage between its minimum and maximum baseline, different from other interferometric arrays such as ASKAP, and a declination-dependent resolution that worsens towards more negative declinations. MOST images of the entire sky below a declination of $-30\degree $ are available from SUMSS, collected at 843 MHz with a narrow, 3 MHz bandwidth. SUMSS data products typically reach a $\sim 1$ mJy/bm RMS sensitivity, across its $43$"$\times43$"$\csc|\delta|$ synthesized beam, making it similar and complementary to the Northern NRAO VLA Sky Survey (NVSS). For this work, we use the SUMSS image mosaic J1130M64, accessed via the University of Sydney repository of mosaics\footnote{\url{http://www.astrop.physics.usyd.edu.au/mosaics/}}.

Secondly, we use ASKAP survey observations taken for the RACS and EMU surveys. RACS targets the sky below a declination of $+41\degree $ to $+49\degree $\citep[depending on the observing frequency;][]{mcconnell2020,racs-mid}, with a higher typical sensitivity and resolution:  $\sim 250$ $\mu$Jy/bm and typical beam sizes between $15$ and $25$ arcseconds. RACS will eventually include data at three frequencies (UHF, L, and S band; RACS-low, mid, and high, respectively); for this work, we access the currently released UHF and L band images of the field surrounding LS 2355 from the CSIRO ASKAP Science Data Archive (CASDA)\footnote{\url{https://data.csiro.au/domain/casdaObservation}}. These observations are taken at centroid frequencies of 887.5 and 1367.5 MHz and bandwidths of 288 and 144 MHz, respectively. We employed the RACS-low field 1135-62A and RACS-mid field 1136-64, both at a common circular beam size of 25 arcseconds.

In the Galactic plane, where LS 2355 is located, both SUMSS and RACS tend to show a higher RMS due to the presence of diffuse sources of emission and complex image artefacts due to bright (extended) structures in the plane. Therefore, we additionally employ ASKAP observations from the EMU survey \citep{norris2011,norris2021}, similarly accessed via the CASDA. The EMU survey, which is ongoing, will perform deep, $10$-hour observations of each field below $+30\degree$ declination. Data for finished fields is publicly available, including the field containing LS 2355. The longer exposure time leads to a lower nominal RMS sensitivity of  $25$-$30$ $\mu$Jy/beam. Data is collected at a centroid frequency of 944 MHz with a bandwidth of 288 MHz, overlapping with the RACS-low band. We specifically accessed the EMU Stokes-I field 1136-64 at high resolution, corresponding to a beamsize of $7.9\times7.3$ {\color{black} arcseconds$^2$} at a position angle of $75.6\degree$. 

In the $\gamma$-ray band, we turn to the latest \textit{Fermi}/Large Area Telescope (LAT) source catalogue, namely its 14-year Source Catalogue 4FGL-DR4\footnote{Accessed via \url{https://fermi.gsfc.nasa.gov/ssc/data/access/lat/14yr\_catalog/}.} \citep{fermisourcecat1,fermisourcecat2}. Compared to the 3FGL catalogue searched by \citetalias{fermipaper2018}, the 4FGL-DR4 catalogue contains significantly longer total exposures, leading to better constrained source positions and spectral measurements. The potential, 3FGL counterpart of LS 2355 identified by \citetalias{fermipaper2018}, 3FGL J1128.7$-$6232, corresponds to the source 4FGL J1130.5$-$6236c in the 4FGL-DR4. This source remains an unidentified \textit{Fermi} source, with strongly improved position accuracy in the new dataset. From the 4FGL-DR4, we know its 90\% positional uncertainty ellipse and spectral properties. 

Finally, to supplement our non-thermal investigation of the LS 2355 bow shock, we also employ IR and optical observations. In the IR band, we use the NASA/IPAC Infrared Science Archive to access \textit{WISE} observations of the field containing LS 2355 from the ALLWISE program\footnote{\url{https://doi.org/10.26131/IRSA153}}. We specifically access images from Band 3 (12.1 micron) and Band 4 (22 micron) for visual comparison with the radio observations. We further quantitatively analyse the Band 3 image, which -- like all ALLWISE images -- is distributed in units of DN/pixel. The typical background value if 500 DN/pixel; when converting these image units to physical flux densities, we use the conversion of $1.83\times10^{-6}$ Jy/DN. In the optical band, we use \textit{Gaia} data from DR3 \citep{gaiamission,gaiadr3} to constrain the movement of LS 2355 (Gaia ID 5333860240705973888) with respect to its surroundings. 

\section{The non-thermal counterpart of the LS 2355 bow shock}

\subsection{The ASKAP detection of the radio bow shock}

The bow shock of LS 2355, not known prior to the work by \citetalias{fermipaper2018}, is clearly visible in infrared and optical images of the surroundings of the runaway star. To assess whether a radio counterpart is present and detectable, we first consider the EMU observations of the field. Figure \ref{fig:emu_full} shows the $1.6\degree\times2\degree$ field of view centred on the position of LS 2355. The Galactic Plane crosses this field from the bottom left to top right; however, while diffuse radio emission may often cause imaging artefacts in such crowded fields, the deep exposures of the EMU survey allow for a relatively artefact-free deconvolution across this specific field. The HII region, GAL 293.60$-$01.28, that LS 2355 is moving towards, is clearly detected, as is substructure within the region: an arc-shaped bow shock that appears radio-brighter than the surrounding HII region, as well as several radio-bright edges towards the Northern edge of GAL 293.60$-$01.28. Based on this deep EMU image, we therefore report the detection of a radio counterpart of the LS 2355 bow shock.

In Figure \ref{fig:emu_full}, we also plot the 90\% positional uncertainty for the proposed \textit{Fermi} counterpart of the bow shock, both based on the 3FGL catalogue \citepalias[following][3FGL J1128.7$-$6232]{fermipaper2018} and the 4FGL-DR4 catalogue (4FGL J1130.5$-$6236c). While the former overlaps in positional uncertainty with both the HII region and the bow shock, the improved positional accuracy in 4FGL-DR4 shows how it likely doesn't overlap with either. The overlap between both error regions is located at the top-left edge of the HII region but does not point towards an obvious radio counterpart, to the depth of the EMU survey. We will discuss the implication of this improved positional uncertainty in 4FGL-DR4 on the non-thermal properties of the LS 2355 bow shock in Section \ref{sec:discussion}. 

\begin{figure}
\includegraphics[width=\columnwidth]{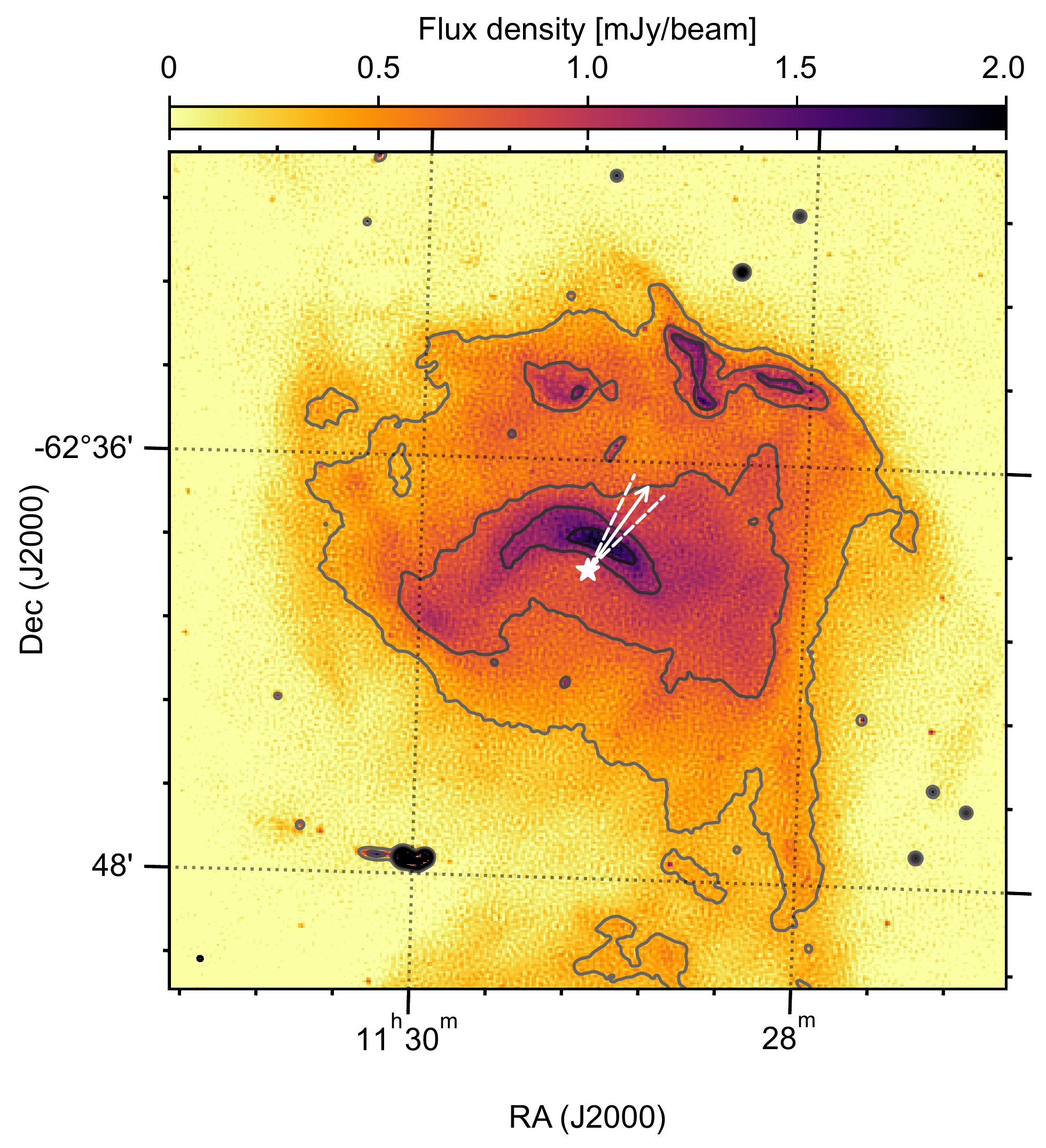}
 \caption{A zoomed-in version of the EMU field shown in Figure \ref{fig:emu_full}, showing the $24\times24$ arcminute$^2$ field around LS 2355. The synthesized beam, with size $7.9\times7.3$ arcseconds$^2$ at a position angle of $75.6\degree$, is shown in the bottom left corner. The arrow indicates {\color{black} the proper motion} of LS 2355, corrected for local Galactic rotation, {\color{black} with the two dashed lines showing the uncertainty on the direction}. The four contour levels shown in this image are plotted at $5$, $10$, $15$, and $20$ times the image RMS sensitivity of $75$ $\mu$Jy/bm. The latter two contours trace the extent and core of the radio bow shock, respectively, while also tracing the radio-bright edges of the HII region towards its top right.}
 \label{fig:emu_zoom}
\end{figure}

To further investigate the EMU radio counterpart of the bow shock, we show a zoomed-in region of the data in Figure \ref{fig:emu_zoom}. In that image, we include four contour levels, at $5$, $10$, $15$, and $20$ times the RMS sensitivity across the field, i.e. $75$ $\mu$Jy/beam. The contours in this image, and all later images, are smoothed with a Gaussian kernel across a smoothing scale of three pixels. The brightest two contours, at $1.125$ mJy/bm and $1.5$ mJy/bm, enclose the ove rall shape and core region of the bow shock. The former of those two levels also constrains the brightest regions of the two radio-bright edge regions of the HII region, towards the top right. We will include these two contours, calculated from the EMU data, in later images of the field to aid comparison. 

\begin{figure}
\includegraphics[width=\columnwidth]{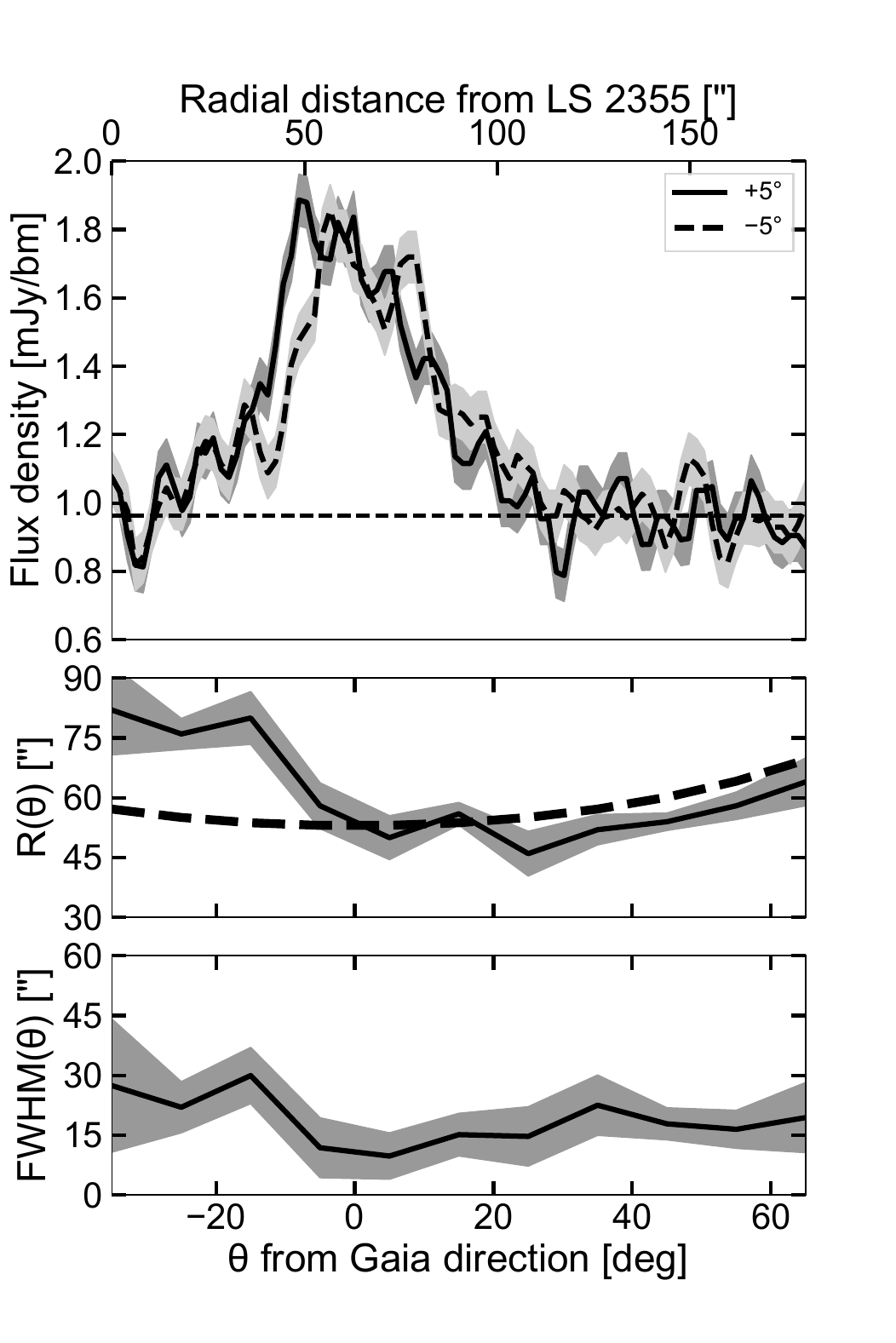}
 \caption{\textit{Top:} the radial flux density profile in the EMU image, {\color{black} along two directions within the uncertainty range of the Galactic-corrected \textit{Gaia} motion of LS 2355 ($\theta = \pm 5\degree)$.} The shaded areas shows the $1\sigma$ uncertainty on the flux density. The dashed horizontal line shows the average level in the final third ($60$ arcseconds) of the profile. From this profile, we measure an angular stand-off distance of $R_0 = 53 \pm 3$ arcseconds. \textit{Middle:} the bow shock radial distance from LS 2355 as a function of the angle $\theta$ (positive equals clockwise) with respect its \textit{Gaia} motion. The dashed line indicates the bow shock shape derived by \citet{wilkin1996}. \textit{Bottom:} the FWHM of the bow shock profile as a function of $\theta$.}
 \label{fig:radial}
\end{figure}

In Figure \ref{fig:radial}, we show the radial and resolved properties of the radio bow shock based on the EMU data. {\color{black} For this purpose, we calculate the profile for directions starting from North ($\theta=0$), increasing clockwise in steps of $10\degree$. In the top panel, we show the flux density profile in two directions within the uncertainty of the movement direction of LS 2355 (see Section \ref{sec:gaia} for its determination based on \textit{Gaia} data). For each profile, we calculate the radial distance $R(\theta)$ corresponding to the peak flux density and its full-width half-maximum $FWHM(\theta)$. For the latter, we calculate the bow shock maximum as the flux density excess above the constant level at radial distances exceeding 2 arcminutes (e.g., the dashed line in the top panel of Figure \ref{fig:radial}).}

From this radial analysis, we measure an angular stand-off distance of {\color{black} $R_0 = 53 \pm 3$ arcseconds, corresponding to $R_0 = 0.57 \pm 0.03$ parsec} at the distance of LS 2355. We can furthermore compare the angular profile $R(\theta)$ with the prediction from \citet{wilkin1996}, plotted in the middle panel of Figure \ref{fig:radial} as the dashed line. While the observed and model shape approximately agree for positive angles $\theta$, larger separations are seen at negative $\theta$. Such asymmetry may result from an inhomogeneous ISM, with a smaller density leading to larger separations. Distortions from the canonical \citet{wilkin1996} shape may also arise from the thermal pressure of the ISM \citep{christie2016,benaglia2021}: while the canonical shape assumes only ISM ram pressure, i.e., a cold ISM, thermal pressure may play a significant role for ionized ISM regions. The ratio between thermal and ram pressure, $r \sim kT_e / m_p v_*^2$, equals $r \sim 0.12$ for $T_e \sim 10^3$ K. An enhanced thermal pressure would not create the observed asymmetric shape. Therefore, the approximate agreement between the observed and model $R(\theta)$ for $\theta > 0$ suggest a relative small $r$; in other words, temperatures not significantly exceeding the $T_e = 10^3$ K mentioned above. We will discuss the effect of thermal pressure in more detail in Section \ref{sec:discussion}.

\subsection{The non-thermal nature of the radio emission}

\begin{figure*}
\includegraphics[width=\textwidth]{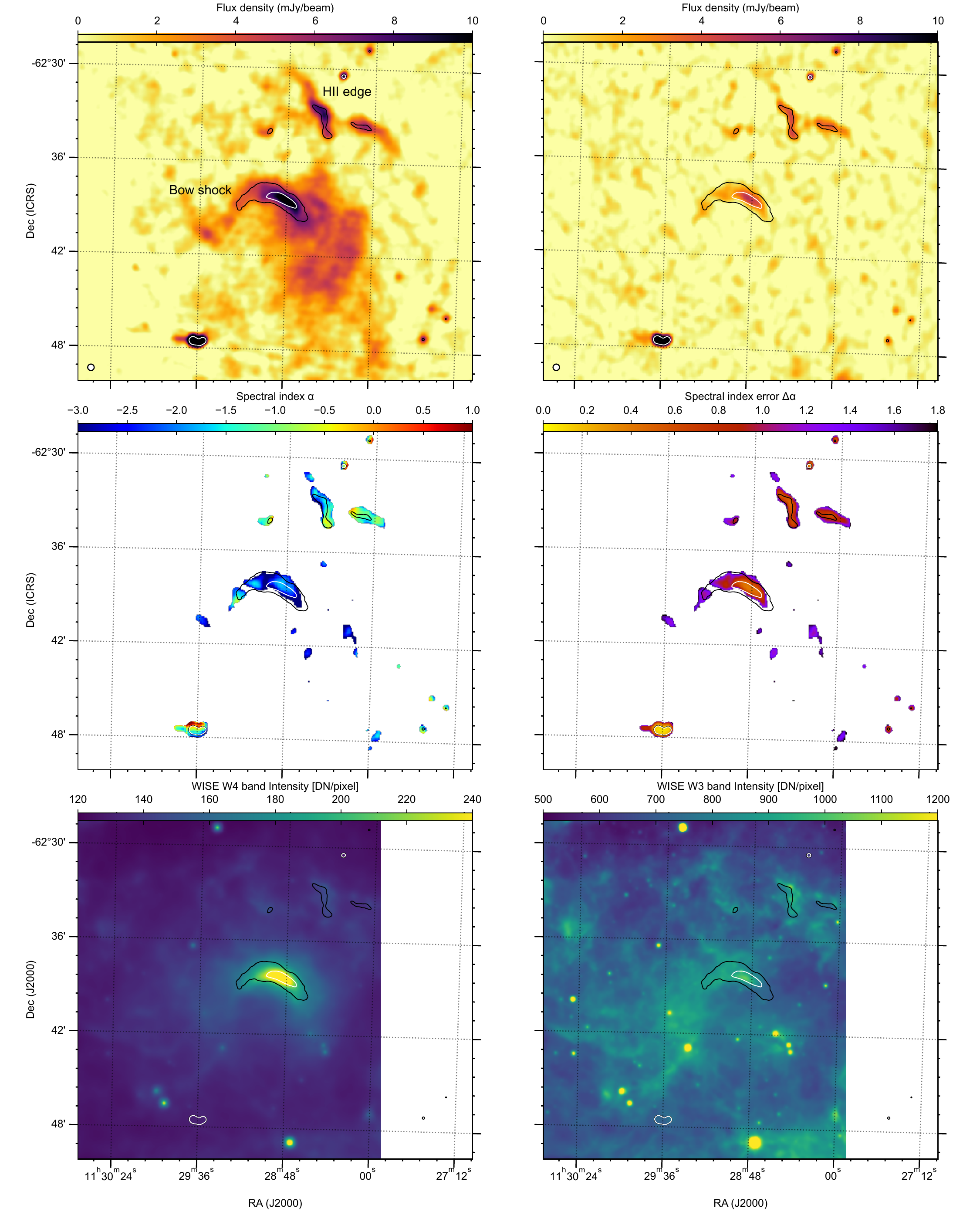}
 \caption{\textit{Top row:} the RACS-Low ({\color{black} left; $500$ $\mu$Jy/bm RMS) and RACS-Mid (right; $300$ $\mu$Jy/bm RMS)} images centred on LS 2355, deconvolved to a common beam size of $25$ arcseconds (shown in the bottom left of both panels). \textit{Middle row:} the radio spectral index (left) and its $1\sigma$ uncertainty, calculated on a per-pixel basis. Only pixels with sufficiently high flux density in both RACS bands are plotted. \textit{Bottom row:} The \textit{WISE} W4 (left) and W3 (right) images. The contours shown in all six panels are the $15\sigma$ and $20\sigma$ contours calculated from the deeper EMU image (see Figure \ref{fig:emu_zoom}). All six panels have a $0.42\degree\times0.36\degree$ size.}
 \label{fig:sixpanels}
\end{figure*}

{\color{black}The EMU observations reveal a radio bow shock counterpart to LS 2355, while the updated Fermi observations show that the originally proposed $\gamma$-ray counterpart is actually offset from both the bow shock and the HII region}. To further investigate the non-thermal properties of the bow shock, we turn to the RACS observations: these survey data include (at the time of writing) two observing frequencies, convolved to a common beam size, making it well suited for spectral index estimates. In the top two panels of Figure \ref{fig:sixpanels}, we show the RACS-Low (left) and RACS-Mid (right) images centred on LS 2355. The plotted contours are the aforementioned $15\sigma$ and $20\sigma$ EMU contours. The bow shock and HII edge regions can be identified in both RACS bands, tracing out the contours, as expected. In the top left panel, we specifically indicate the HII edge  region that we will apply below in our evaluation of the bow shock spectral index. The bottom two panels show the \textit{WISE} W4 (left) and W3 (right) bands, where the bow shock predominantly shows up in the lower resolution W4 band. In the RACS images, the shock is marginally resolved in the radial direction, given their common, $25$ arcsecond beam size (leading to poorer radial resolving power compared to the EMU resolution; e.g. Figure \ref{fig:radial}). The larger beam size also results in a high peak flux density compared to EMU (cf. the colorbar scaling in Figure \ref{fig:emu_zoom}), despite their overlapping frequency bands. 

In the middle panels of Figure \ref{fig:sixpanels}, we plot the spectral index $\alpha$ (left) and its $1\sigma$ uncertainty $\Delta \alpha$ (right), where its sign is defined according to flux density $S_\nu \propto \nu^\alpha$.  The spectral index is calculated for each pixel. Despite the same beam size, the two RACS datasets have slightly different pixel sizes: the spectral index map is calculated at the lowest-resolution pixel size of the two. We apply a flux density threshold in both bands before calculating the spectral index: $\alpha$ is only calculated for pixels where $S_{\rm low} \geq 2.5$ mJy/beam and $S_{\rm mid} \geq 1$ mJy/beam, {\color{black} corresponding to $5$ and $\sim3.3$ times the image RMS, respectively; these values are optimized such that only pixels with reasonably low uncertainty are shown}. The uncertainty on $\alpha$ is then calculated using its definition and error propagation. 

Resulting from the flux density thresholding, mainly the three aforementioned regions within the HII region -- the bow shock and the two edge regions -- appear in the spectral index map. Both appear with negative spectral indices, although the bow shock shows significantly steeper values than the HII edge regions. These values are artificially steepened, as discussed below: the average value within the $20\sigma$ bow shock contour is $\alpha_{\rm bowshock} = -2.2$ with an average uncertainty $\Delta \alpha_{\rm bowshock} = 0.5$, compared to average values within the $15\sigma$ contour in the largest of the two HII edge regions of $\alpha_{\rm HII-edge} = -1.4$ and $\Delta\alpha_{\rm HII-edge} = 0.5$. Such artificial steepening of the spectrum is expected based on the frequency difference and the fixed array configuration: with fixed baselines, the RACS-Mid images will resolve out and lose more extended flux than the RACS-Low images. {\color{black} We note that the uncertainty quoted above, and plotted in Figure \ref{fig:sixpanels}, does not include any systematic uncertainty on the flux; adding a $1$\% systematic flux scaling uncertainty (see Section \ref{sec:33}) increases both average uncertainty levels to $\Delta \alpha_{\rm stat+syst} = 0.9$.}

\begin{figure}
\includegraphics[width=\columnwidth]{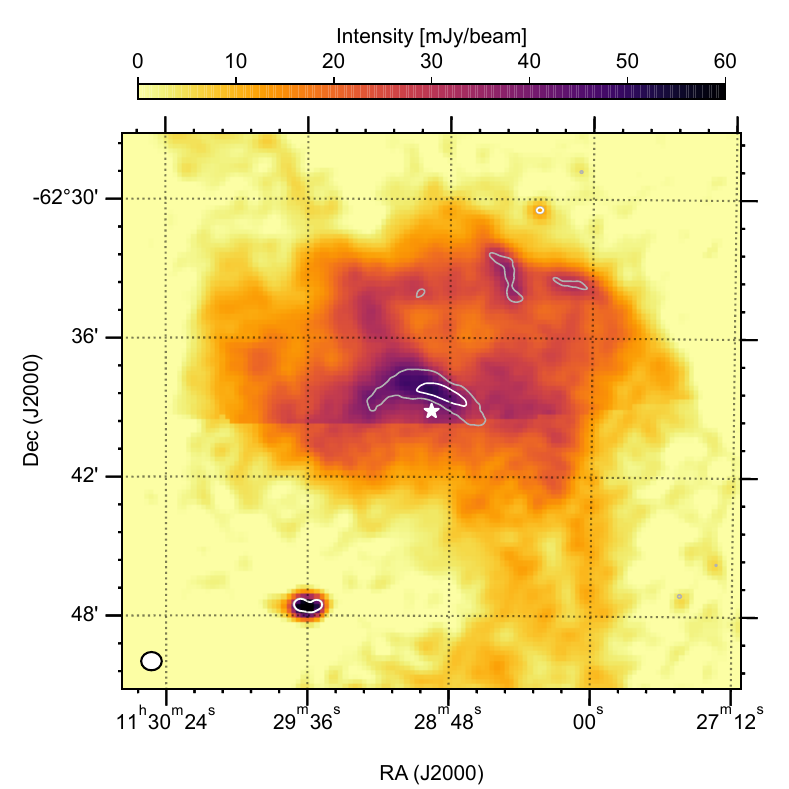}
 \caption{The SUMSS image of the region around LS 2355, indicated by the white star. The plotted contours are the $15\sigma$ and $20\sigma$ contours calculated from the deeper EMU image (see Figure \ref{fig:emu_zoom}). The beam size is shown in the bottom left panel; a mosaicing artefact can be identified in the image as the roughly horizontal line across the middle of the image.}
 \label{fig:SUMSS}
\end{figure}

To assess the level to which emission is resolved out, a Total Power measurement with a single dish telescope can be employed. However, given the complex field, such a measurement for the LS 2355 bow shock would likely not distinguish between the different sub-structures within the HII region. Instead, we turn to SUMSS, which overlaps in frequency with RACS-Low and offers continuous uv-plane coverage down to significantly smaller baselines than ASKAP. As a result, it provides a balance between the need for resolution to resolve the bow shock and the need for a Total Power measurement. Figure \ref{fig:SUMSS} shows the SUMSS survey data, again including the EMU contours. A comparison between RACS-Low and SUMSS shows that for both the bow shock and the largest HII edge region, a similar level of emission is resolved out by ASKAP: the total flux density within the $20\sigma$ bow shock contour is $58.3$ mJy and $40.8$ mJy in SUMSS and RACS-Low, respectively; for the HII edge region, these values within the $15\sigma$ contour are $38.1$ mJy and $25.4$ mJy, respectively. RACS-Low therefore recovers $70\%$ and $67\%$ of the SUMSS flux density in the bow shock and edge regions, respectively -- a similarity that is unsurprising, given the similar size and orientation of the two compared regions. 

SUMSS only covers a single frequency band and can therefore not be used to obtain a similar correction in L band (RACS-Mid). However, the analysis above indicates that similar levels of flux density are lost in the HII edge region and the bow shock: a conclusion that is expected to apply similarly in L band. If we assume that the RACS-Low and RACS-Mid flux densities, across region of similar angular size and orientation, are a fraction $f_{\rm low}$ and $f_{\rm mid}$, respectively, of the intrinsic flux densities $S_{\rm UHF}$ and $S_{\rm L}$, we can relate the observed to the intrinsic spectral index via:

\begin{align}
    \alpha = \frac{\log\left(S_{\rm mid} / S_{\rm low}\right)}{\log\left(\nu_{\rm mid}/\nu_{\rm low}\right)} &= \frac{\log\left(S_{\rm UHF} / S_{\rm L}\right)}{\log\left(\nu_{\rm mid}/\nu_{\rm low}\right)} + \frac{\log\left(f_{\rm mid} / f_{\rm low}\right)}{\log\left(\nu_{\rm mid}/\nu_{\rm low}\right)} \nonumber \\
 &= \alpha_{\rm intrinsic} + C \text{ where } C < 0 \text{,}
\end{align}

\noindent where the last line follows from $f_{\rm mid} < f_{\rm low}$ while $\nu_{\rm mid} > \nu_{\rm low}$. Importantly, the spectral index offset caused by resolving out increasingly more emission at higher frequencies, is \textit{constant} -- regardless of the actual flux density levels of different regions. As we have observed that the bow shock and the HII edge region lose similar levels of flux between MOST and ASKAP at UHF band, we will assume that both regions undergo the same \textit{constant} shift in spectral index. 

Under this assumption, we therefore conclude that the bow shock has a steeper spectrum than the edge of the HII region, making it the steepest spectrum region within the total HII region. If we assume that the edges of HII region are caused by brightenings of their thermal emission, we expect them to display an instrinsic spectral index of $\alpha_{\rm intrinsic} = -0.1$. That interpretation would imply $C \approx -1.3$ and an average intrinsic spectral index across the bow shock region of $\alpha_{\rm bowshock} = -0.9$ with an average uncertainty of {\color{black}$\Delta \alpha = 0.7$ (statistical only; increasing to $\Delta \alpha = 1.2$ when including the systematic uncertainty)}. Here, we stress that these values are based on the average pixel values of $\alpha$, despite their spatial variation. {\color{black} If the nature of the HII edge region is different, and for instance contains non-thermal contributions from shock acceleration by expansion of the HII region \citep{padovani2019,dewangan2020}, the offset $C$ would be different; in that scenario, the intrinsic bow shock spectral index would be steeper. Combined, the average uncertainty and the effect of this thermal assumption imply that the exact spectral index of the bow shock remains challenging to constrain. However,} both the qualitative conclusion regarding its steep spectrum and the quantitative estimates of $\alpha_{\rm bowshock}$ point towards a non-thermal nature for the radio bow shock of LS 2355.

\subsection{The multi-wavelength observational properties of the bow shock and LS 2355}
\label{sec:33}

From these radio (and infrared) images, we can not only identify the bow shock counterpart and assess its non-thermal nature, but also measure several key observables. All properties that we derive and list below, are also summarized in Tables \ref{tab:surveys} and \ref{tab:input}; the former table discusses the observational properties of the four used radio surveys, while the latter table contains the derived measurements from the radio, IR, and optical data.

Firstly, for our later calculations, it is essential to measure the total radio flux density across the bow shock. As our modelling work will employ a single-zone model, we adopt the $20\sigma$ EMU survey contour that traces the core region of the bow shock well in both radio and IR.  While this region only encapsulates the brightest, central regions of the radio counterpart, it will suffice for the modelling performed in the next sections. This same contour-based definition will also be applied for the other measured quantities of the bow shock. Across this region, the integrated EMU flux density is $S_\nu = 79.4\pm0.8$ mJy, where the uncertainty is calculated as follows: {\color{black} for the statistical error ($0.3$ mJy), we use the error propagation, by multiplying the image RMS calculated in a source-free region of the image (see Table \ref{tab:surveys}), with the square root of the number of beams covered by the region. We combine this error in quadrature with an assumed $1$\% systematic flux density uncertainty to account for absolute flux calibration.} In Table \ref{tab:input}, for completeness, we also list the integrated flux density of the $15\sigma$ EMU bow shock contour, as well as the integrated flux density of the bright edge of the HII region within its $15\sigma$ contour. Finally, the Table also lists the average spectral indices and average spectral index errors measured for the bow shock and edge region, as well as the index offset assuming the edge region emits optically thin thermal radio emission. 

The ASKAP EMU bow shock data can also be expressed in brightness temperature following 
\begin{equation}
    T_{\rm B} = 1.22\times10^3 \left(\frac{I}{\text{mJy/bm}}\right) \left(\frac{\nu_{\rm obs}}{\text{GHz}} \right)^{-2} \left(\frac{\theta_{\rm maj}\theta_{\rm min}}{\text{arcsec}^2} \right)^{-1} \text{K .}
\end{equation}
\noindent Brightness temperature, when sufficiently large, can further support a non-thermal over a thermal interpretation of the emission. For the EMU survey however, where the bow shock peaks at approximately $3$ mJy/bm, the brightness temperature is $T_{\rm B} \sim 70$ K; in other words, the brightness temperature is too low to rule out thermal emission and therefore independently distinguish between the two potential emission mechanisms. 

In addition, several geometrical measurements will be used as input for our later calculations. We again consider the central bow shock area enclosed by the $20\sigma$ contour in the EMU data, to remain consistent with the above flux density measurements. Its area is $3.1\times10^3$ arcsec$^2$; as the width of the bow shock region at its apex we measure $\Delta \sim 32$ arcseconds. We assume a distance to both LS 2355 and the bow shock equal to the \textit{Gaia} eDR3 distance to LS 2355, $D=2.2\pm0.1$ kpc \citep[][]{Bailer-Jones2021}, which is consistent with but better constrained than the value used by \citetalias{fermipaper2018}. Assuming that the depth of the bow shock is of similar order to its width, we can express its width, area, and volume in physical units as $\Delta = 0.35$ pc, $A_{\rm BS} = 0.35$ pc$^2$, and $V_{\rm BS} = 0.12$ pc$^3$, respectively. For the standoff distance between the shock and LS 2355, we adopt the aforementioned value measured from the $R(\theta=\pm5\degree)$ profiles, which is consistent with infrared and optical constraints \citepalias{fermipaper2018}. Finally, the volume factor, as defined in \citet{vandeneijnden2022_velx1} to capture the fraction of a sphere with radius $R_0$ overlapping with the bow shock, equal $\eta_{\rm vol} = 0.114$. 

\label{sec:wisedata}
The \textit{WISE} infrared data allows us to estimate the dust temperature and infrared photon density, which are both necessary input for the calculations of the inverse Compton scattering processes in the bow shock \citep{delvalle2012,debecker2017}. Across the considered bow shock area, the {\color{black} \textit{WISE} Band 4 image contains a summed value of $5.9\times10^5$ in the pixel units of DN. The typical background across the image is $\sim 120$ DN/pixel, contributing a total of $2.8\times10^5$ DN in the bow shock region. The net shock flux density can be calculated using the \textit{WISE} Band 3 conversion of $5.23\times10^{-5}$ Jy/DN, which implies an integrated $22$ micron flux density of $F_{\rm IR} \approx 16.2\pm0.3$ Jy. The uncertainty on the IR flux density is calculated in the same manner as the radio flux density, scaling the RMS sensitivity in DN/pixel across a source free region ($\sim 120$ DN/pixel) to the bow shock area, and applying the DN to Jy correction.} For the luminosity of LS 2355, which in turn affects the dust temperature, we adopt the measurement from \citet{hohle2010}: $L_{\rm Bol} = 2\times10^4$ $L_{\odot} = 0.77\times10^{38}$ erg/s.

\label{sec:gaia}
In their original analysis, \citetalias{fermipaper2018} calculated the proper motion of LS 2355 based on six position measurements spanning roughly a century. Their analysis, performed shortly after the launch of \textit{Gaia}, does not use \textit{Gaia} data beyond its first position measurement. For our work, we make use of the third \textit{Gaia} data release, which contains accurate proper motion measurements for LS 2355. Its proper motion in equatorial coordinates $(\alpha, \delta)$ is measured as $\mu_{\alpha*} \equiv \mu_{\alpha} \cos\delta = -6.41 \pm 0.02$ mas/yr and $\mu_\delta = 1.68 \pm 0.02$ mas/yr, which represents a significantly smaller motion in declination than measured by \citet{fermipaper2018}\footnote{\citet{fermipaper2018} already note, using only the first \textit{Gaia} position, an apparent deviation in the \textit{Gaia} declination from their historic trend.}.  We then follow the prescription in \citet{comeron2007} and convert this proper motion to Galactic coordinates, before applying their Equations 2a and 2b to calculate the movement of the ISM local to LS2355 and subtracting it from the \textit{Gaia} proper motion. Finally, we convert the corrected Galactic proper motion back to the equatorial frame. {\color{black} In the calculation of the local movement, we use $(U,V,W)_\odot = (11.1, 12.24, 7.25)$ km/s \citep{schonrich2010}. When using the same Oort's constants as in \citet{comeron2007}, i.e. $A = -B = 12.5$ km/s/kpc, we find $\mu_{\alpha*\rm, corr}  = -0.61 \pm 0.02$ mas/yr and $\mu_{\delta\rm, corr} = 0.52 \pm 0.02$ mas/yr, implying an angle of $49\degree$ West of North.  

However, while the statistical uncertainty from the \textit{Gaia} data implies small uncertainties on this direction, and the stellar speed, the assumption regarding the Oort's constants yields larger systematic uncertainties: assuming a $1$ km/s/kpc uncertainty on both constants implies a $18\degree$ and $3$ km/s uncertainty on direction and speed, respectively. Using more recent estimates of the Oort's constants from \citet{bovy2017}, with $A = 15.3 \pm 0.4$ km/s/kpc and $B = -11.9 \pm 0.4$ km/s/kpc, we instead find $\mu_{\alpha*\rm, corr}  = -0.34 \pm 0.14$ mas/yr and $\mu_{\delta\rm, corr} = 0.45 \pm 0.05$ mas/yr. These values imply an angle of $35\pm10\degree$ West of North and a speed of $6.0\pm1.2$ km/s. Given that the latter direction aligns better with the apex of the bow shock, we adopt that as the direction plotted in Figure \ref{fig:emu_zoom} (where we also show the uncertainty). {\color{black} To calculate the stellar speed, we similarly use the Oort's constants from \citet{bovy2017}, as well as the radial velocity of LS 2355 as reported in Appendix A of \citetalias{fermipaper2018}. Specifically, in this calculation, we convert this reported heliocentric radial velocity to the local radial velocity. The resulting, total stellar velocity, at the assumed distance of $2.2$ kpc, is $v_* = 7.0 \pm 2.5$ km/s, substantially lower than the $v_* = 23$ km/s found by \citetalias{fermipaper2018}. We explicitly address the effect that this value of the velocity has on our later calculations, and the discrepancy with the substantially different value, $v_* = 23$ km/s, found by \citetalias{fermipaper2018}, in Section \ref{sec:discussion}.}

\begin{table}
\caption{Observational details of the four consulted radio surveys in this work. See Section \ref{sec:data} for details and references.}
\label{tab:surveys}
\begin{tabular}{llllll}
\hline
\textit{} & Quantity [unit] & EMU & \multicolumn{2}{c}{RACS} & SUMSS \\
\textit{} & \textit{} & \textit{} & \textit{Low} & \textit{Mid} & \textit{} \\

\hline
$\nu_{\rm obs}$ & Frequency [MHz] & 944 & 887.5 & 1367.5 & 843 \\
$\Delta \nu$ & Bandwidth [MHz] & 288 & 288 & 144 & 3 \\
$S_{\rm RMS}$ & RMS [$\mu$Jy/bm] & 75 & 500 & 300 & 3000 \\
$\theta_{\rm maj}$ & Major beam  ["] & 7.9 & 25 & 25 & 50.7 \\
$\theta_{\rm min}$ & Minor beam ["] & 7.3 & 25 & 25 & 43.0 \\
BPA & Position angle [$\degree$] & 75.6 & -- & -- & 0.0 \\
\hline

\end{tabular}\\
\end{table}

\begin{table*}
\caption{The parameters of LS 2355 and its bow shock, observed or derived in this work and used in the calculations in Section \ref{sec:discussion}. The final column notes what data source underlies the measurement, including the region definition, if relevant.}
\label{tab:input}
\begin{tabular}{llll}
\hline
Parameter & Quantity & Value & Reference / dataset \\ \hline
$S_{\rm bowshock}$ & Flux density of bow shock at $944$ MHz & $79.4 \pm 0.8$ mJy & EMU (within $20\sigma$ contour) \\ %
$S_{\rm bowshock}$ & Flux density of bow shock at $944$ MHz & $358.9\pm3.7$ mJy & EMU (within $15\sigma$ contour) \\ %
$S_{\rm HII-edge}$ & Flux density of HII edge at $944$ MHz & $53.8 \pm 0.6$ mJy & EMU (within $15\sigma$ contour) \\ 
\hline
$\alpha_{\rm bowshock}$ & Mean spectral index in bow shock & $-2.2$ & RACS (within $20\sigma$ EMU contour) \\
$\Delta\alpha_{\rm bowshock}$ & Mean spectral index error in bow shock & {\color{black} $0.5$ (stat) / $0.9$ (stat + syst)} & RACS (within $20\sigma$ EMU contour)  \\
$\alpha_{\rm HII-edge}$ & Mean spectral index in HII edge region & $-1.4$ & RACS (within $15\sigma$ EMU contour)  \\
$\Delta\alpha_{\rm HII-edge}$ & Mean spectral index error in HII edge region & {\color{black} $0.5$ (stat) / $0.9$ (stat + syst)} & RACS (within $15\sigma$ EMU contour) \\
 &  & $0.9$ (stat + syst) & \\
$C$ & Spectral index offset if $\Delta\alpha_{\rm HII-edge} \equiv -0.1$ & $-1.3$ & RACS / SUMSS (within $15\sigma$ EMU contour) \\
\hline
$D$ & Distance & $2.2\pm0.1$ kpc & \citet{Bailer-Jones2021} \\ %
$R_0$ & Standoff distance & $0.57 \pm 0.03$ pc & EMU (radial profile) \\ %
$\Delta$ & Bow shock width & $32$ arcsec = $0.35$ pc & EMU ($20\sigma$ contour) \\ %
$A_{\rm bowshock}$ & Bow shock surface & $3.1\times10^3$ arcsec$^2$ = $0.35$ pc$^2$ & EMU ($20\sigma$ contour) \\ %
$V_{\rm bowshock}$ & Bow shock volume & $9.9\times10^4$ arcsec$^3$ = $0.12$ pc$^3$ & EMU ($20\sigma$ contour) \\ %
$\eta_{\rm vol}$ & Volume factor & $0.114$ & EMU ($20\sigma$ contour) \\ \hline
$\mu_{\alpha} \cos\delta$& \textit{Gaia} proper motion (RA) & $-6.41 \pm 0.02$ mas/yr & \citet{gaiadr3} \\
$\mu_\delta$ & \textit{Gaia} proper motion (Dec) & $1.68 \pm 0.02$ mas/yr & \citet{gaiadr3} \\
$(\mu_{\alpha} \cos \delta)_{\rm, corr}$ & Corrected proper motion (RA) & {\color{black} $-0.34 \pm 0.14$ mas/yr} & This work \\
$\mu_{\delta\rm, corr}$ & Corrected proper motion (Dec) & {\color{black} $0.45 \pm 0.05$ mas/yr} & This work \\
$v_*$ & LS 2355 velocity& {\color{black} $7.0\pm2.5$ km/s} & This work, based on \citet{gaiadr3}\\ 
$L_{\rm bol}$ & LS 2355 bolometric luminosity & $0.77\times10^{38}$ erg/s & \citet{hohle2010} \\
\hline %
$F_{\rm IR}$ & Infrared flux density$^{*}$ & $16.2 \pm 0.3$ Jy & ALLWISE -- Band W4 (within $20\sigma$ EMU contour) \\
$\nu_{\rm IR}$ & \textit{WISE} infrared observing frequency & $1.38\times10^{13}$ Hz & ALLWISE -- Band W3 (within $20\sigma$ EMU contour) \\
$T_{\rm dust}$ & Bow shock dust temperature & $41$ K & \citet{hohle2010} \\
$a_{\rm dust}$ & Dust grain size & $0.2$ $\mu$m & \citet{draine1981}; \citet{delvalle2012} \\
\hline
\end{tabular}\\
\end{table*}

\section{Discussion}
\label{sec:discussion}

In this paper, we present the detection of non-thermal radio emission from the bow shock of LS 2355. Re-analysing \textit{Fermi} catalogue data, we also conclude that the potential $\gamma$-ray counterpart proposed by \citetalias{fermipaper2018}, is an unrelated source: it's enhanced position is significantly offset from both the bow shock and the HII region that LS 2355 is moving towards. Finally, we employ \textit{Gaia} data on LS 2355 to update its proper motion, finding a lower velocity with respect to its surroundings compared to earlier estimates.  We measure a central bow shock flux density of $79.4 \pm 0.8$ mJy at 944 MHz, using ASKAP observations from the EMU survey. This flux density corresponds to a radio luminosity of $L_R = \nu F_\nu = 4\times10^{29}$ erg/s. Assuming a volume factor $\eta_{\rm vol} = 3V_{\rm bow shock}/ 4\pi R_0^3 = 0.114$, this luminosity corresponds to a fraction $\eta_{\rm radio} \approx 10^{-5} (\dot{M}_{\rm wind} / 10^{-6} \text{ } M_\odot/\text{yr})^{-1} (v_{\infty}/1000\text{ km/s})^{-2}$ of the available kinetic wind power, where $\dot{M}_{\rm wind}$ and $v_{\infty}$ are the mass-loss rate and terminal velocity of the stellar wind launched by LS 2355, respectively. 

These findings make LS 2355 the third runaway massive star with a spectrally-confirmed non-thermal bow shock. The first example, BD +43$\degree$ 3654, was reported by \citet{benaglia2010} and later re-observed and re-analysed by \citet{benaglia2021}, \citet{moutzouri2022}, and \citet{martinez2023}. Broadband radio observations indicate a spectral index of approximately $\alpha \approx -1$, although \citet{martinez2023} stress that resolved-out emission at higher frequencies may somewhat artificially steepen the spectrum. The second source, BD +60$\degree$ 2522 was reported by \citet{moutzouri2022}, with a similarly steep radio spectrum ($\alpha \approx -0.8$). Finally, \citet{martinez2023} argue that the radio bow shocks of G1, G3, and Vela X-1 \citep{vandeneijnden2022_velx1,vandeneijnden2022_racs} may be dominated by or contain a significant contribution of non-thermal emission. However, currently, all three lack published radio spectral constraints. 

The three sources with detected non-thermal emission, as well as G1, G3, and Vela X-1, are all located at relatively close-by distances and have ordinary, relatively similar mass-loss properties. The main difference between LS 2355 and the five other sources lies in its proper motion and ISM surroundings. It moves more slowly than all other systems (a factor $\sim 2$ slower than G1, the slowest of those five), but encounters a dense and highly-structured ISM: where all six targets are located in relatively complex ISM regions, as shown by the presence of other extended radio sources, LS 2355 interacts directly with a dense HII region (GAL 293.60$-$01.28).  Again scaling with typical stellar wind parameters, the encountered ISM density can be written as $n_{\rm ISM} = 1.3\times10^2 \text{ } (\dot{M}_{\rm wind} / 10^{-6} \text{ } M_\odot/\text{yr})(v_{\infty}/1000\text{ km/s})$ (where we ignore thermal pressure for this order of magnitude scaling). Such values are consistent with the expectations for an HII region but are easily one to two orders of magnitude larger than those seen in the other five sources. It is the low stellar velocity, implying a smaller ISM ram pressure, that causes the bow shock to be observed at detectable offset from the star nonetheless.

Beyond the basic estimates above, we can further investigate the non-thermal properties of the bow shock with a simplified, one-zone approach. As discussed by \citet{martinez2023}, and later in this section, this simplified analytical approach is less accurate than multi-zone modelling. We limit this work to the former method, but will discuss the potential limits to our inference, leaving the latter approach to future work. The basic physical scenario follows the commonly proposed setup for non-thermal bow shock emission \citep{delvalle2012,delpalacio2018,delvalle2018,martinez2023}: the stellar wind provides a kinetic reservoir that powers the acceleration of particles at the shock through diffusive shock acceleration. The resulting population of accelerated relativistic electrons may then lose energy via radiative processes, importantly synchrotron emission in the presence of the shock's magnetic field and inverse Compton scattering interactions with the ambient infrared photon field from dust and stellar emission. Alternatively, particles may leave the acceleration region via advective or diffusive escape. The synchrotron process is responsible for the observed radio emission; the inverse Compton scattering dominates at high energies, where, as we conclude in this work, no $\gamma$-ray (or X-ray) counterpart is detected.

We first turn to the magnetic field present in the bow shock, by assessing the equipartition magnetic field and the maximum field strength. The former is the field strength where the combination of energy stored in the accelerated particle population (probed by $S_\nu$) and in the magnetic field is optimized; the latter is the magnetic field where the magnetic pressure equals the thermal pressure. For larger magnetic field strengths, and therefore magnetic pressures, the material becomes incompressible, preventing the formation of the shock and resulting diffusive shock acceleration. 

\begin{figure}
\includegraphics[width=\columnwidth]{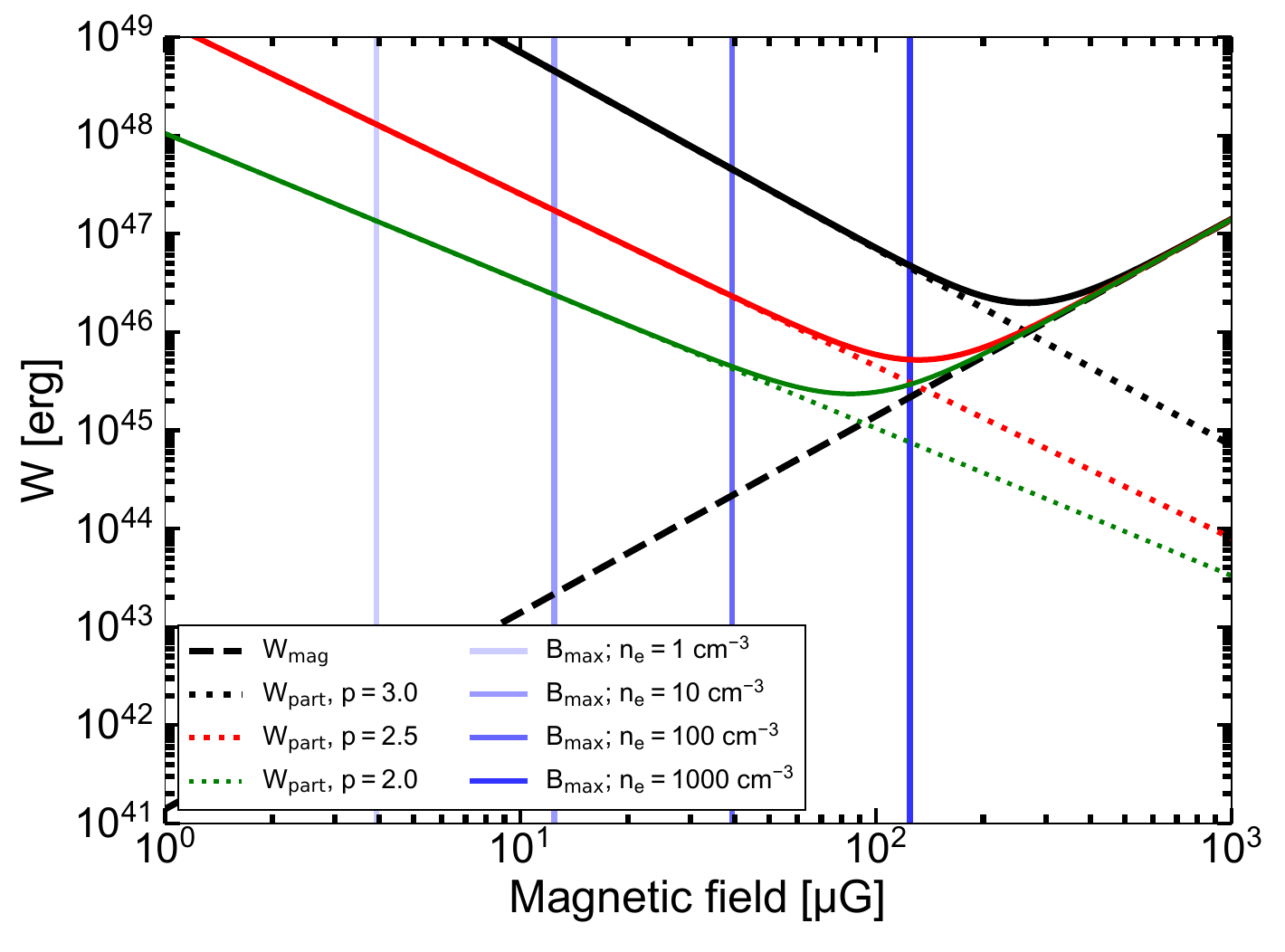}
 \caption{The energy contained in relativistic particles $W_{\rm part}$ and the magnetic field $W_{\rm mag}$, as a function of the magnetic field strength, for three different values of $p$. We also show the maximum magnetic field for four different electron densities of the ISM.}
 \label{fig:eq}
\end{figure}

\begin{figure*}
\includegraphics[width=\textwidth]{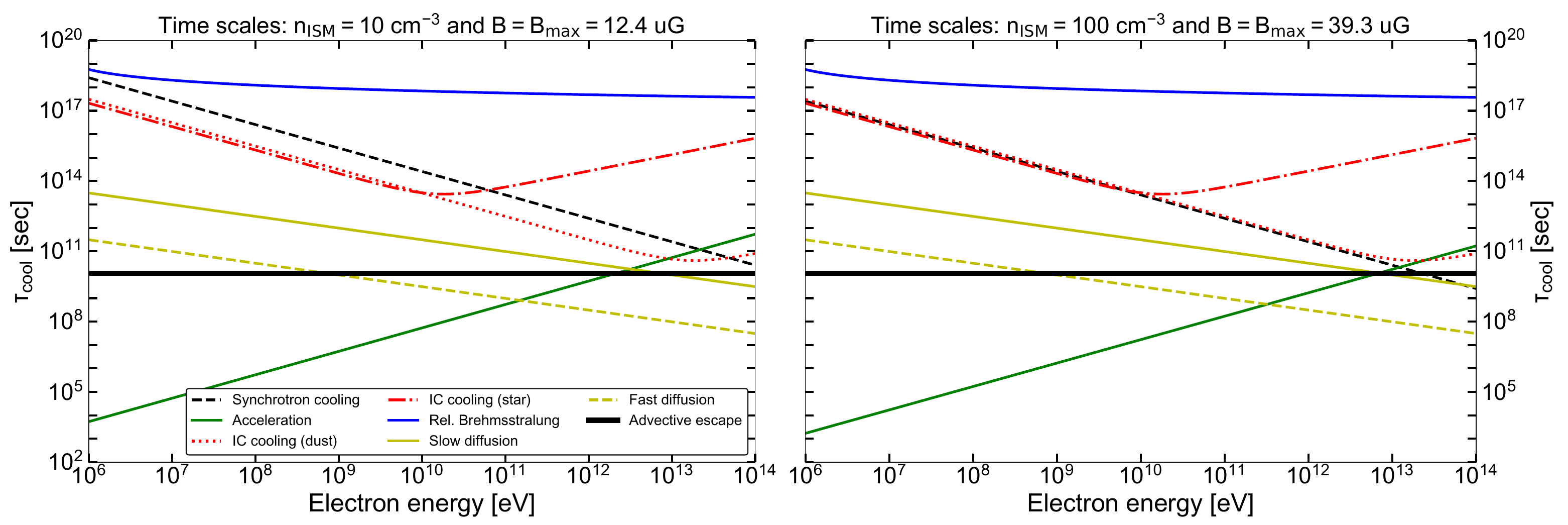}
\caption{The time scales of radiative and dynamic electron cooling in the shock, assuming two ISM densities and the resulting maximum magnetic field. The different curves show the time scales of escape due to synchrotron losses, inverse Compton losses, relativistic Brehmsstrahlung, diffusion, and advective escape, as a function of electron energy. The time scale of acceleration is shown as well; the maximum particle energy is constrained by finding the energy where the particle acceleration time scale first equals a loss time scale.}
\label{fig:tau}
\end{figure*}

For the equipartition analysis, we follow standard practice\footnote{As outlined in e.g. \citet{longair2011} and expressed in Equation A1 to A5 of \citet{vandeneijnden2022_velx1}.}, where the electron number density distribution takes a power law form, $N(E) = \kappa E^{-p}$, between an energy $E_{\rm min}$ and $E_{\rm max}$. Specifically, we assume the electrons to be relativistic ({\color{black} $E_{\rm min} = 1$ MeV}), reaching a maximum energy of $E_{\rm max} = 10^{13}$ eV (see below; note that these equipartition inferences are not significantly affected by the exact value of $E_{\rm max}$). We consider three values of $p$: {\color{black} $p=2$, $p=2.5$, and $p=3$. These values are chosen to cover the expected range for the bow shock, as the effect of resolved out emission and the uncertainty on the spectral index mean that a single value is challenging to select. The three values correspond to a range of spectral indices between $\alpha = -0.5$ and $\alpha = -1$.} We plot the particle and magnetic energy as a function of magnetic field in Figure \ref{fig:eq}. For the three plotted cases, the equipartition magnetic field is substantial, of the order $B_{\rm eq} \gtrsim 100$ $\mu$G. 

The maximum magnetic field can be estimated via \citep{delpalacio2018,benaglia2021} 
\begin{equation}
    \frac{B^2}{2\mu_0} \leq \frac{2}{1+\gamma_{\rm ad}} \rho_{\rm wind} v^2_{\infty} \text{ ,}
\end{equation}
\noindent where the right-hand side represents the thermal pressure. With $\gamma_{\rm ad} = 5/3$ and the definition for the wind density and stand-off distance, this equation is equivalent to $B_{\rm max} = \sqrt{1.5\mu_0 m_p n_{\rm ISM} v^2_*}$. Barring direct measurement of the ISM density of the HII region, we indicate the maximum field for four values logarithmically spaced from $1$ to $10^3$ cm$^{-3}$ in Figure \ref{fig:eq}.

Interestingly, this comparison shows that the maximum magnetic field strength is lower than the equipartition field for most considered ISM densities: only for the highest considered density in combination with shallow electron energy distributions, the two become comparable. Therefore, in the remainder of this discussion, we will assume that the system is out of equipartition and close to its maximum magnetic field; in practical terms, that $B=B_{\rm max}$ for the considered ISM density. This analysis also indicates that the ISM density is likely amongst the higher values considered here: not only are those values consistent with an HII region, lower values imply significantly larger total particle energies at $B_{\rm max}$, requiring significantly higher acceleration efficiency. We finally note that this magnetic field strength may be scaled towards a stellar magnetic field for LS 2355, as $B_{*} \approx 2\times10^2 (n_e / 10^2\text{ cm}^{-3}) (R_{*}/10R_{\odot})^{-1}$ G \citep{delpalacio2018}; a value consistent with measurements in populations of O-type stars \citep{rustem2023}.

Building on the above analysis, i.e. assuming $B=B_{\rm max}$, we can consider the relevant time scales at play in the bow shock. At its core, we follow the analysis presented in \citet{vandeneijnden2022_velx1}. For this calculation, we assume that the time scale of advective escape from the shock can be estimated as the bow shock width $\Delta$ divided by the stellar wind velocity; for simplicity, we assume a single velocity of $1000$ km/s, but note that the deviations in this value are expected to be significantly smaller than the many orders of magnitude spanned by the time scales of the different considered processes. In addition to the existing analyses of \citetalias{fermipaper2018} (for the LS 2355 bow shock, specifically) and \citet{vandeneijnden2022_velx1}, we include diffusive escape following the parameterization of \citet{delvalle2018}. For this purpose, we assume that the energy-dependent diffusive escape occurs on a time scale of 

\begin{equation}
\tau_{\rm diffusion} \equiv \frac{R_0^2}{D(E)} = \frac{R_0^2}{D_{10}} \left(\frac{E_{\rm electron}}{10\text{ GeV}}\right)^{-\delta_{\rm diff}} \text { .}
\end{equation}

\noindent Here, we follow \citet{delvalle2018} and assume that $\delta_{\rm diff} = 0.5$. We similarly consider a case of slow and fast diffusion, as captured by the constant $D_{10}$: $D_{10} = 10^{25}$ cm$^2$/s for slow diffusion, and $D_{10} = 10^{27}$ cm$^2$/s for fast diffusion. The latter value follows the fast diffusion scenario in \citet{delvalle2018}.

In Figure \ref{fig:tau},  we show the results of this exercise for two cases, defined by a different ISM density. This density, as discussed above, affects the assumed magnetic field, e.g., $B_{\rm max}$, which in turns affects the acceleration and synchrotron time scales. Similarly, the ISM density sets the mass-loss rate, as we assume a wind velocity of $10^3$ km/s, and therefore the cooling time scale of relativistic Brehmsstralung. The left-hand and right-hand panels show the case of $n = 10$ cm$^{-3}$ and $n=100$ cm$^{-3}$, respectively. In both scenarios, slow diffusion and advective escape place similar limits to the particle energy; in the latter case, the higher magnetic field and therefore faster synchrotron losses imply that those also place a similar limit. In either scenario, a maximum energy in the range $10^{12} - 10^{13}$ eV is expected, unless fast diffusion is operating in the bow shock.  

The approach of the analytic estimates above treats the entire central bow shock region as a single object. In their recent modelling work, \citet{martinez2023} point out that such single-zone modelling can lead to different conclusions than more involved multi-zone modelling that includes the resolved structure of the bow shock. In particular, single-zone modelling with a single electron energy power-law distribution may underestimate the contribution of non-thermal emission, leading to an overestimate of the required efficiency of electron acceleration. For example, where \citet{vandeneijnden2022_velx1} use a single-zone approach to conclude that the radio emission from Vela X-1 (where no spectral shape has been measured) is dominated by thermal emission, \citet{martinez2023} instead conclude that, despite a thermal contribution, non-thermal emission is the dominant factor. Both approaches, however, appear to be liable to overestimating the thermal contribution: even in sources observed with a non-thermal radio spectrum (BD +43$\degree$ 3654 and BD +60$\degree$ 2522), \citet{martinez2023} find that the expected thermal contribution overpredicts the observed flux densities. Such overestimation is likely the result of Kelvin Helmholtz instabilities \citep{comeron1998} that are not included in current modelling. The inclusion of a scaling factor, $\eta_H < 1$, in the width of the isothermal layer of the shocked ISM, reducing its resulting thermal emission, is used to account for this issue by \citet{martinez2023}.

\begin{figure*}
\includegraphics[width=\textwidth]{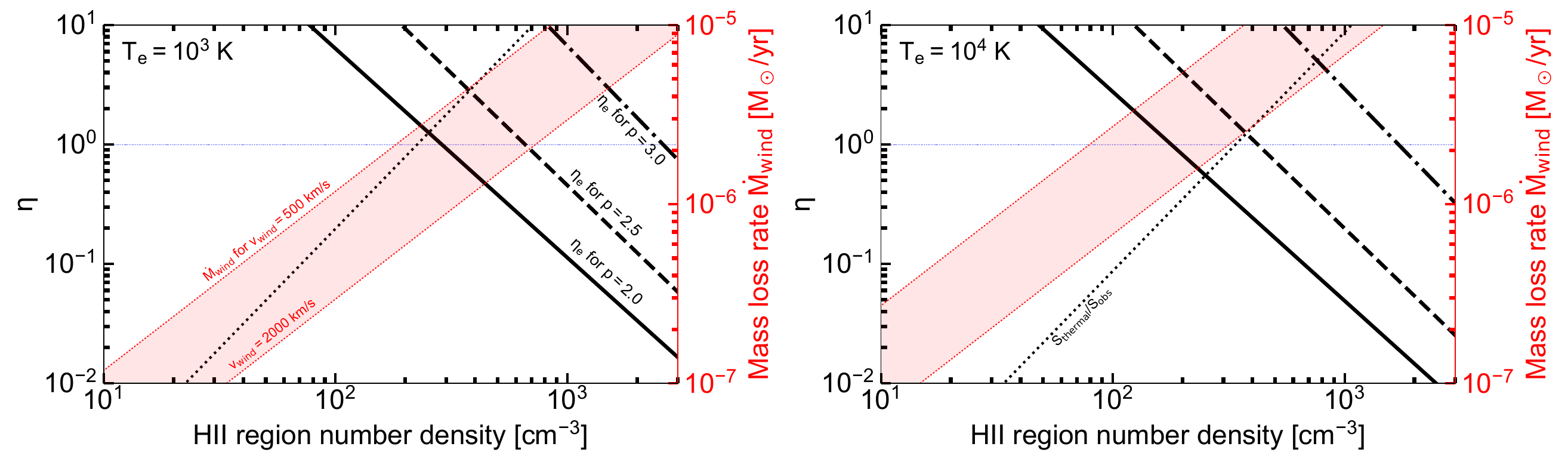}
\caption{The comparison of thermal emission, non-thermal efficiency, and stellar wind parameters, for different ISM properties. All three quantities are plotted as a function of the number density of the ISM, specifically of the HII region that LS 2355 is moving into. The temperature of the ISM, affecting the thermal emission directly, as well as the thermal pressure of the ISM -- stellar wind interaction, differs between the panel: $T_e = 10^3$ K and $T_e = 10^4$ K, in the left-hand and right-hand panel, respectively. \textit{Left-hand axis:} the efficiency of particle acceleration $\eta_e$ for three values of p (2.0, 2.5, and 3.0 for the line, dashed line, and dash-dotted line, respectively) and the ratio $\eta_{\rm th}$ between predicted thermal radio flux density $S_{\rm thermal}$ and the observed flux density $S_\nu$ at 944 MHz (dotted line). \textit{Right-hand axis:} the mass-loss rate of LS 2355 for a range of wind velocities between 500 and 2000 km/s.  The thin, horizontal dotted line in both panels indicates $\eta = 1$ to guide the eye.}
\label{fig:parspace}
\end{figure*}

We find that these two known effects -- an underestimation of the non-thermal contribution when treating the central shock as one zone and an excessive thermal contribution -- may also play a role for the LS 2355 bow shock. We display this visually in Figure \ref{fig:parspace}, where we plot three quantities as a function of assumed density of the ISM, for two ISM temperatures. Plotting the dependence of these three quantities on just density, for two temperatures, allows us to consider whether a reasonable segment of parameter space exists that explains the observed radio data. The three quantities we plot are: the efficiency of particle acceleration as defined in \citet{vandeneijnden2022_velx1}; the expected thermal emission of the shocked ISM $S_{\rm thermal}$, as a fraction of the observed emission $S_\nu$ (both plotted using the same left-hand axis); and the wind mass-loss rate for a range of wind velocities (right-hand axis). The efficiency of particle acceleration, $\eta_e$, depends on ISM density via both the magnetic field, for which we again assume $B=B_{\rm max}$, and the stellar wind properties. As was discussed earlier in this section, $n_{\rm ISM}$ depends on the mass-loss rate and terminal velocity, which implies the $\dot{M}_{\rm wind} v_{\infty}$ factor in the efficiency can be calculated for a given ISM density. When we properly include the thermal pressure for a given ISM temperature, we find following \citet{martinez2023} that $\dot{M}_{\rm wind} v_\infty = 4 \pi n_{\rm ISM} R_0^2 (m_p v_*^2 + kT_e)$. We apply this approach for the stellar wind parameters in the denominator of the acceleration efficiency equation from \citet{vandeneijnden2022_velx1}:

\begin{equation}
\begin{split}
        \eta_e = &\frac{128\pi^3 R_0^3 D^2 S_\nu \epsilon_0 c m_e}{3\sqrt3 \dot{M}_{\rm wind} v_\infty \Delta V_{\rm bowshock} e^3 B_{\rm max} a(p)}\\
    &\times \left(\frac{3eB_{\rm max}}{2\pi\nu m_e^3 c^4}\right)^{-(p-1)/2} \int_{E_{\rm min}}^{E_{\rm max}} E^{1-p}dE \text{ .}
    \end{split}
    \label{eq:radioonly_NT}
\end{equation}

Figure \ref{fig:parspace} shows that large ISM densities are required to yield an acceleration efficiency substantially below unity; a qualitatively expected trend, given that higher densities imply both a higher maximum magnetic field and larger stellar wind pressure, and therefore wind power budget. We show the same relations for an ISM temperature of $T_e = 10^3$ K (left) and $T_e = 10^4$ K (right). In the former case with relatively shallow electron energy distributions  -- {\color{black} the canonical $p=2$, specifically} -- efficiencies around $10\%$ are seen for densities around $10^3$ cm$^{-3}$; for $p=3.0$, {\color{black} the steepest value plotted here}, the required densities are so high as to fall beyond the plotted scale. The former densities may be consistent with an HII region, and the corresponding mass-loss rates are similarly feasible for LS 2355: $6\times10^{-6}$ $M_\odot$/yr for a terminal velocity of $1000$ km/s. As discussed above, however, a challenge is posed by the thermal emission, which significantly over-predicts the total flux density at such ISM densities. As a significant thermal contribution is not consistent with the observed non-thermal spectrum, a significant reduction of the thermal contribution would be required: $\eta_H \ll 1$, in the terminology introduced by \citet{martinez2023}. In the case of higher ISM temperature (right panel), these inferences change slightly: for shallower spectra, densities or $\sim (5-7)\times10^2$ cm$^{-3}$ are sufficient for a $10$\% efficiency {\color{black} in the case of a shallow spectrum ($p=2$)}. However, at these densities, a significant thermal contribution is similary expected -- $\eta_H \ll 1$ remains required. While the enhanced thermal pressure may increase the required wind power generally, the reduction in ISM density largely counters that: $\dot{M}_{\rm wind} \approx 7\times10^{-6}$ $M_\odot$/yr for a terminal velocity of $1000$ km/s and $n_{\rm ISM} = 5\times10^2$ cm$^{-3}$. 

{\color{black} In the context of the above discussion, where the ISM particle density is taken as the independent unknown variable, we can briefly consider whether free-free absorption further constrains the realistic parameter space. Free-free absorption at radio frequencies, i.e. a free-free optical depth $\tau_{\rm ff} \gtrsim 1$, leads to a optically thick, strongly inverted spectrum with $\alpha = +2$. The edge of the HII region does not show evidence for such a spectrum. Following Equation 5 in \citet{bloot2022} and assuming that the depth of the HII edge is similar to its width (i.e. of the order of the beam size of 25 arcseconds, or $\sim 0.27$ pc in physical units), we estimate an optical depth of $\tau_{\rm ff} \approx 0.1$ at 1 GHz for $n_e \sim 10^3$ cm$^{-3}$ and $T_e = 10^4$ K. This value, which increases towards lower temperatures and quadratically increases with number density, implies free-free absorption does not play a dominant role in the considered parameter space for $T_e = 10^4$ K. However, it also shows that densities exceeding $n_e \sim 10^3$ cm$^{-3}$ by a factor of a few are unlikely, as absorption would come into play -- barring significant changes in density across the HII region.}

These inferences, implying relatively large required acceleration efficiencies and a correction to the thermal contribution for both considered ISM temperatures, can also be viewed through a different lens: for instance, the large required ISM densities that lead to bright expected thermal emission, are driven by the low velocity inferred from \textit{Gaia}: at a higher velocity, the ISM -- stellar wind ram pressure balance requires lower densities. In our analysis, we correct the observed proper motion for the approximate motion of objects on a circular Galactic orbit at the distance and Galactic position of LS 2355. If this approximation causes an underestimated stellar velocity relative to its direct surroundings, its effect on the expected thermal emission is substantial: for a given stand-off distance, $n_{\rm ISM} \propto v_*^{-2}$ (ignoring thermal pressure), while $S_{\rm thermal} \propto n_{\rm ISM}^2$. The used {\color{black} $v_* = 7.0\pm2.5$ km/s} is only marginally super-sonic for a surrounding isothermal ISM at $T \approx 10^4$ K; the velocity inferred by \citetalias{fermipaper2018}, $v_* = 23$ km/s, would change the expected relative thermal contributions by a {\color{black} factor $\sim 1\times10^{-2}$.}

While a full multi-zone modelling effort is beyond the scope of this observationally-focused paper, such modelling will be vital to assess the particle acceleration efficiency and magnetic field morphology of the shock in more detail. To better understand the effect of resolved-out emission on the observed spectral index, forward modelling of the radio observations is a key further step: here, it is vital to account for the detailed uv-plane coverage of the considered observations, therefore taking into account the specific observatory, array configuration, observing setup, and source position and elevation. On the observational side, a wider range of covered frequencies will provide better constraints on the non-thermal spectral shape and the underlying electron population -- in particular the overall shape beyond a single power-law model without an exponential cutoff at high energies \citep{martinez2022,martinez2023}. For LS 2355, observations at S band with MeerKAT, or in the future with SKA-mid and SKA-low, would provide such extended frequency coverage. 

Finally, at the end of this Discussion, we briefly return to the originally-proposed $\gamma$-ray association by \citetalias{fermipaper2018}. While the updated \textit{Fermi} FGL4-DR4 catalogue data rules out an associated on spatial grounds, our analysis shows that this association would similarly be challenged by the energetics of the system. The low magnetic field strengths inferred by \citetalias{fermipaper2018} ($B < 1$ $\mu$G) would imply significantly larger total power budgets; as can be seen in Figure \ref{fig:eq}, such fields imply an enhancement in particle energy of at least two orders of magnitude, implying an extreme required stellar wind power budget. Alternatively, the significantly larger magnetic fields inferred in our work would imply, as shown in Figure \ref{fig:tau}, maximum electron energies inconsistent with the $\gamma$-ray spectral turnover observed in the originally-proposed \textit{Fermi} counterpart; only an unexpectedly fast diffusion, beyond the fastest scenario plotted in Figure \ref{fig:tau}, would sufficiently reduce the expected maximum energy. The lack of a $\gamma$-ray counterpart to a non-thermal radio-bright bow shock is also consistent with the prediction by \citet{delpalacio2018}, i.e., that radio-bright systems are not necessarily the best $\gamma$-ray targets. However, the inverse dependence of their predicted radio-to-$\gamma$-ray luminosity ratio on ISM density, could imply that the high ISM densities surrounding LS 2355 may be an interesting target for future, targeted $\gamma$-ray follow up; more extensive modelling of this bow shock system is required to further substantiate such expectations.

\section{Conclusions}

In this work, we have reported an in-depth radio study of the bow shock of LS 2355. We explore observations from the ASKAP and MOST telescopes to identify the radio counterpart of the bow shock and the larger-scale environment with which LS 2355 is interacting. Using multi-band RACS data, we infer that the radio emission from the bow shock is of non-thermal nature, making it the third example of a spectrally-confirmed non-thermal bow shock driven by a massive runaway star. To investigate the potential association of this bow shock with an unidentified \textit{Fermi} 3FGL $\gamma$-ray source, we search the updated 4FGL-DR4 catalogue for counterparts in these deeper data. While the originally-proposed counterpart is present in the updated catalogue, its improved positional accuracy argues strongly against its association with the bow shock. Finally, we update the proper motion and stellar velocity measurements of LS 2355 using \textit{Gaia}. Our initial single-zone analytical modelling suggests the system resides in sub-equipartition with a magnetic field likely close to the maximum field strength allowed by the balance of magnetic and thermal ISM pressures. It further implies that LS 2355 interacts with an ISM with substantial density, as expected for an HII region. Such densities suggest a thermal emission contribution from the shocked ISM that is substantially larger than what the non-thermal nature of the radio spectrum allows; the presence of currently unmodelled instabilities in the shocked ISM and a potential underestimate of the stellar velocity may alleviate this issue. 

\section{Acknowledgements}
For the purpose of open access, the authors have applied a Creative Commons Attribution (CC-BY) licence to any Author Accepted Manuscript version arising from this submission. The authors thank the referee, whose constructive review greatly improved the scope and depth of this work. The authors also thank Tara Murphy, Laura Driessen, and Kovi Rose for useful discussions on ASKAP and SUMSS observations and spectral index estimates. JvdE acknowledges a Warwick Astrophysics prize post-doctoral fellowship made possible thanks to a generous philanthropic donation. FC acknowledges support from the Royal Society through the Newton International Fellowship programme (NIF/R1/211296). This research has made use of NASA's Astrophysics Data System Bibliographic Services. This scientific work uses data obtained from Inyarrimanha Ilgari Bundara / the Murchison Radio-astronomy Observatory. We acknowledge the Wajarri Yamaji People as the Traditional Owners and native title holders of the Observatory site. CSIRO’s ASKAP radio telescope is part of the Australia Telescope National Facility (\url{https://ror.org/05qajvd42}). Operation of ASKAP is funded by the Australian Government with support from the National Collaborative Research Infrastructure Strategy. ASKAP uses the resources of the Pawsey Supercomputing Research Centre. Establishment of ASKAP, Inyarrimanha Ilgari Bundara, the CSIRO Murchison Radio-astronomy Observatory and the Pawsey Supercomputing Research Centre are initiatives of the Australian Government, with support from the Government of Western Australia and the Science and Industry Endowment Fund. This paper includes archived data obtained through the CSIRO ASKAP Science Data Archive, CASDA (\url{http://data.csiro.au}). This research has made use of the NASA/IPAC Infrared Science Archive, which is funded by the National Aeronautics and Space Administration and operated by the California Institute of Technology. This work has made use of data from the European Space Agency (ESA) mission \textit{Gaia} (\url{https://www.cosmos.esa.int/gaia}), processed by the \textit{Gaia} Data Processing and Analysis Consortium (DPAC, \url{https://www.cosmos.esa.int/web/gaia/dpac/consortium}). Funding for the DPAC has been provided by national institutions, in particular the institutions participating in the \textit{Gaia} Multilateral Agreement.

\section*{Data Availability}

All observational data used in the paper is available publicly in the data repositories of the respective observatories and surveys. A \textsc{GitHub} reproduction repository for all analysis and calculations in this paper is available at \url{https://github.com/jvandeneijnden/LS2355}. This repository includes \textsc{Jupyter} notebooks performing all calculations underlying this paper, as well as reproducing all figures. The repository therefore also includes the \textsc{fits} images of the analysed data from SUMSS, RACS, and \textit{WISE}. To ensure long-term reproducibility, the repository is also be available as a stable release via Zenodo, linked via the GitHub page.

\bibliographystyle{mnras} 
\bibliography{references.bib}

%%%%%%%%%%%%%%%%%%%%%%%%%%%%%%%%%%%%%%%%%%%%%%%%%%

% Don't change these lines
\bsp	% typesetting comment
\label{lastpage}
\end{document}